\documentclass[twocolumn, superscriptaddress, showpacs, amsmath, amssymb, aps, prab]{revtex4-1}

\draft 
\usepackage{amsmath}
\usepackage{xcolor}

\usepackage{graphicx}
\usepackage{hyperref}
\usepackage{placeins} 
\begin{document}

\newcommand{\blue}[1]{\textcolor{blue}{#1}}
\newcommand{\red}[1]{\textcolor{red}{#1}}


\title[Sample title]{Self-stabilizing positron acceleration in a plasma column}

\author{S.~Diederichs}
 \email{severin.diederichs@desy.de.}
\affiliation{Deutsches Elektronen-Synchrotron DESY, Notkestr. 85, 22607 Hamburg, Germany}
\affiliation{Lawrence Berkeley National Laboratory, 1 Cyclotron Rd, Berkeley, California 94720, USA}
\affiliation{University of Hamburg, Institute of Experimental Physics, Luruper Chaussee 149, 22607 Hamburg, Germany}

\author{C.~Benedetti}
\affiliation{Lawrence Berkeley National Laboratory, 1 Cyclotron Rd, Berkeley, California 94720, USA}%
\author{M.~Thévenet}
\affiliation{Deutsches Elektronen-Synchrotron DESY, Notkestr. 85, 22607 Hamburg, Germany}
\author{E.~Esarey}
\affiliation{Lawrence Berkeley National Laboratory, 1 Cyclotron Rd, Berkeley, California 94720, USA}%
 
\author{J.~Osterhoff}

\affiliation{Deutsches Elektronen-Synchrotron DESY, Notkestr. 85, 22607 Hamburg, Germany}

\author{C.\,B.~Schroeder}
\affiliation{Lawrence Berkeley National Laboratory, 1 Cyclotron Rd, Berkeley, California 94720, USA}
\affiliation{Department of Nuclear Engineering, University of California, Berkeley, California 94720, USA}

\date{\today}

\begin{abstract}
Plasma accelerators sustain extreme field gradients, and potentially enable future compact linear colliders. 
Although tremendous progress has been achieved in accelerating electron beams in a plasma accelerator, positron acceleration with collider-relevant parameters is challenging. 
A recently proposed positron acceleration scheme relying on the wake generated by an electron drive beam in a plasma column has been shown to be able to accelerate positron witness beams with low emittance and low energy spread. 
However, since this scheme relies on cylindrical symmetry, it is possibly prone to transverse instabilities that could lead, ultimately, to beam break-up. 
In this article, we show that the witness beam itself is subject to various damping mechanisms and, therefore, this positron acceleration scheme is inherently stable towards misalignment of the drive and witness beams. This enables stable, high-quality plasma-based positron acceleration.
\end{abstract}


\maketitle 

\section{Introduction}
The ability to build the next, TeV-class linear particle collider is severely constrained  due to their high construction costs and power consumption. Providing extreme accelerating gradients, plasma-based accelerators potentially enable compact linear colliders, promising drastic cost reductions~\cite{Schroeder:2010,Joshi:2020}. In a plasma-based accelerator, either an intense laser pulse \cite{Esarey:2009} or an ultra-relativistic, high-charge particle bunch \cite{Chen:1985} drives a plasma wake, which can sustain accelerating gradients on the order of tens to hundreds of GV/m. While high-energy-gain~\cite{Blumenfeld:2007,Gonsalves:2019}, high-efficiency~\cite{Litos:2014}, and low-energy-spread~\cite{Lindstrom:2021} electron acceleration was demonstrated experimentally, positron acceleration with collider-relevant parameters is significantly more challenging.

Several concepts have been proposed, including utilizing positron drive beams~\cite{Corde:2015}, use of hollow-core drive beams~\cite{Jain:2015} or lasers pulses~\cite{Vieira:2014b}, a combination of  particle and laser drivers~\cite{Reichwein:2022}, or use of the rear portion of a blowout bubble wake~\cite{Lotov:2007}. Unfortunately, these concepts either lack low emittance, low energy spread, or reasonable efficiency. Hollow core plasma channels have been a promising candidate~\cite{Schroeder:1999, Gessner:NatCom:2016}, but they suffer from intrinsic instability due to the absence of focusing fields for the drive beam~\cite{Schroeder:1999, Lindstrom:2018}. Using asymmetric drive beams provides stability in at least one transverse direction~\cite{Zhou:2021}, but only positron beams with large beam emittances ($>50\,\mu $m\,rad) could be accelerated, which is too large for collider applications. It was found that the wake generated in a thin, warm, quasi-hollow plasma channel 
provides accelerating fields for positrons while being robust against instabilities~\cite{Silva:2021}; however, this scheme was demonstrated for a positron beam with several $\mu $m\,rad emittance and several percent relative energy spread only, i.e., beam quality too poor for a collider. 

In Ref.~\cite{Diederichs:2019}, a concept was proposed that utilizes an electron drive beam and a plasma column to generate positron-accelerating and focusing wakefield structures, and has shown sub-$\mu$m\,rad emittance and sub-percent energy spread positron acceleration \cite{Diederichs:2019, Diederichs:2020}. Since the scheme relies on cylindrical symmetry, one might expect it to be prone to beam breakup instabilities similar to the ones affecting the hollow core plasma channel. In a recent study~\cite{Diederichs:2022:DBS}, the electron driver was found to propagate stably in a plasma column for initial misalignments smaller than the beam size. Still, neither the effect of a misaligned drive beam on the witness beam, nor the stability of a misaligned witness beam itself have been investigated so far.

In this article, we demonstrate by means of theory and particle-in-cell (PIC) simulations the stability of a positron witness beam in the plasma column configuration in the case of a misaligned drive beam or when the witness beam itself is misaligned, and we discuss the corresponding witness beam quality deterioration in presence of such asymmetries.
We show that the witness beam is subject to various damping mechanisms and, therefore, this positron acceleration scheme is inherently stable with respect to misalignment of both the drive and witness beams.
These results pave the path for stable acceleration of low-emittance, low-energy-spread positron beams, a critical step towards the realization of a plasma-based electron-positron collider.

The article is organized as follows: In Sec.~\ref{sec:pos_acc_plasma_column}, we recapitulate positron acceleration in a plasma column. In Sec.~\ref{sec:stability_witness}, the stability of the positron beam in the presence of initial misalignments is investigated. 
In Sec.~\ref{sec:effect_misalignment}, the effect of initial misalignments on the positron beam quality (e.g., deterioration of emittance, energy spread) is determined. Sec.~\ref{sec:conclusion} concludes this work.

\section{Positron acceleration in a plasma column} \label{sec:pos_acc_plasma_column}

\begin{figure}
	\centering
	\includegraphics[trim={0 0 0 0},clip, width=3.375in]{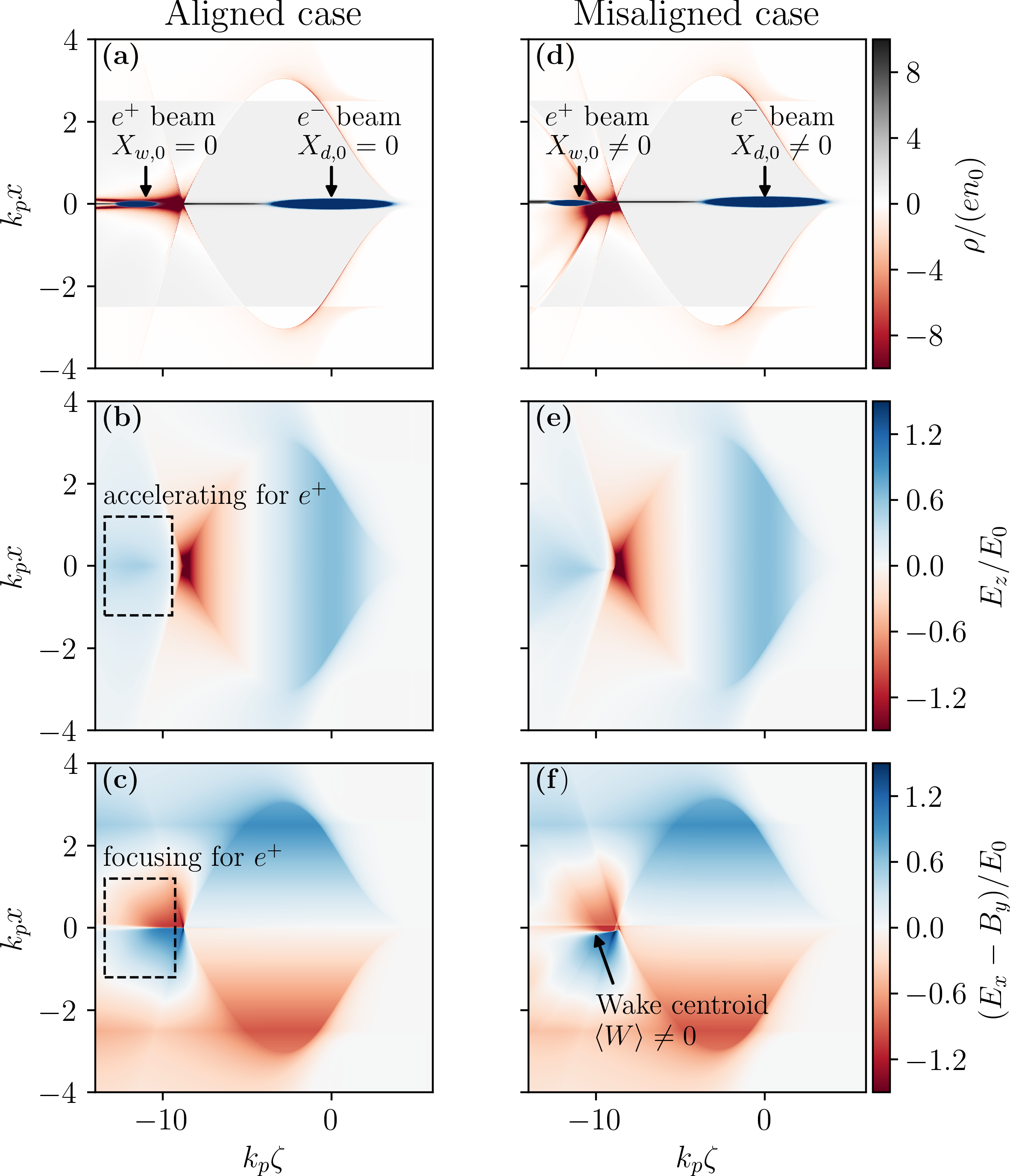}
	\caption{Wakefield generated in a plasma column by an aligned (left column) and misaligned (right column) drive beam. 
	The normalized plasma charge density, the accelerating field, and the focusing field in the $x$-$\zeta$-plane are shown for the case with aligned drive and witness beams (a)--(c), and for misaligned drive and witness beams with offsets of $X_{d,0} = X_{w,0} = 0.05 \,k_p^{-1}$ (d)--(f). As shown in (f), an offset of the drive beam of $X_{d,0} >0$ results in a wake centroid of $\langle W \rangle < 0$. \label{fig:scheme} } 
\end{figure}

As was first discussed in Ref.~\cite{Diederichs:2019}, if an electron beam drives a wake in a plasma column with a column radius smaller than the blowout radius, the transverse wakefields are reduced outside the column due to the lack of ions. The reduced focusing fields induce a spread in the background plasma electron trajectories moving near the boundary of the blowout wake, which return to the propagation axis over an elongated longitudinal region, forming a high-density electron filament at some distance behind the drive beam. The filament creates a wakefield suitable for acceleration and transport of a trailing witness positron bunch, as shown on the left column (a)--(c) of Fig.~\ref{fig:scheme}, where we plot the plasma charge density (a), normalized to $e n_0$, where $e$ is the elementary charge and $n_0$ the ambient plasma density, respectively, and the accelerating (b) and focusing (c) fields, normalized to the cold, non-relativistic wave-breaking field $E_0 = m_e c^2 k_p / e$, in the $x$-$\zeta$-plane. Hereby, $k_p = (4\pi n_0 e^2/m_e c^2)^{1/2}$ is the plasma wavenumber, $e$ the electron charge, $m_e$ the electron mass, $c$ the speed of light in vacuum, $x$ the transverse coordinate, $\zeta = z - ct$ the longitudinal co-moving variable, with $z$ and $t$ being the longitudinal coordinate and the time, respectively. We use a bi-Gaussian electron drive beam with a peak current $I_{d}/I_A = 1$, and  root-mean-square (rms) sizes $\sigma_{z,d} = 1.41\,k_p^{-1}$ (longitudinal) and $\sigma_{x,d} = \sigma_{y,d}= 0.05\,k_p^{-1}$ (transverse), with $I_A = m_e c^3/e \simeq 17$ kA being the Alfv\'en current. The plasma column has a radius $R_p = 2.5\,k_p^{-1}$. In the presented configuration, a helium plasma ionized to the first level is assumed within the column radius and neutral helium gas outside of the column radius. Helium is optimal owing to its high ionization threshold. The usable density range for the plasma column scheme is limited by the wakefield-induced ionization at the boundary of the column. Since the wakefield amplitude scales as $E \sim E_0 \propto n_0^{1/2}$, if the density is too high, the wakefield within the blowout radius (which exceeds the column radius) ionizes the neutral gas outside of the column, causing the column to expand and perturb the positron accelerating and focusing fields. For the above drive beam parameters and a helium plasma, the maximum density is $\simeq 5\times 10^{17}$\,cm$^{-3}$ before field ionization occurs outside the column.

A transverse displacement of the drive beam centroid by one rms size, $X_{d,0} = \sigma_{x,d}$, modifies the electron trajectories asymmetrically, perturbing the wake centroid $\langle W \rangle$ at the back of the blowout. The effect of a misaligned drive beam is shown on the right column (d)--(f) of Fig.~\ref{fig:scheme}. The stable propagation of a misaligned electron drive beam has been shown in a recent study~\cite{Diederichs:2022:DBS}. There, the transversely displaced drive beam was found to be attracted to the center of the plasma column. Although initial misalignment was found to seed the hosing instability~\cite{Whittum:1991}, the induced oscillation is damped by various well-known mechanisms such as ion motion~\cite{Mehrling:PRL:2018}, energy spread~\cite{Mehrling:PRL:2017}, or others~\cite{MartinezDeLaOssa:2018,Lehe:PRL:2017}. Thus, transversely displaced drive beams undergo damped oscillations and drift towards the plasma column axis. The effect of a misaligned drive beam on the positron witness bunch is discussed in the next sections.

\section{Stability of the witness beam} \label{sec:stability_witness}
Based on the premise that the drive beam is stable, we now investigate the witness beam stability and the witness beam quality degradation in the presence of a misaligned drive or witness beam using 3D PIC simulations with the quasi-static code HiPACE\texttt{++}~\cite{Diederichs:2022:HiPACE}. In the simulations, the computational domain is $(-16, 16) \times (-16,16) \times (-14, 6) \, k_p^{-3}$ in $x \times y \times \zeta$, where $x$ and $y$ are the transverse coordinates. The mesh resolution is $0.0078 \times 0.0078 \times 0.001\,k_p^{-3}$. The same drive beam and helium plasma column parameters as in Sec.~\ref{sec:pos_acc_plasma_column} are used. The helium plasma is modeled by electrons and ions with 16 macro-particles per cell each. The drive beam has an initial energy of $5\,$GeV, no energy spread, and a normalized emittance of $\epsilon_0 = 0.18\,k_p^{-1}$, which is matched to the focusing field provided by the background ions in the column. The drive beam is sampled with $10^7$ macro-particles. The simulation is propagated for 1000 time steps with a constant time step of $\Delta_t = 5 \, (c k_p)^{-1}$. 

The witness beam has a transversely Gaussian and a longitudinally tailored profile. To preclude hosing suppression due to the presence of a slice-dependent energy chirp along the positron beam~\cite{Mehrling:PRL:2017} (the energy spread required to suppress the instability is of several percent, and this might be incompatible with collider applications),
we use a witness beam current profile that optimally loads the wake, minimizing the development of a correlated energy spread.
As shown in Ref.~\cite{Diederichs:2020}, the current profile that flattens the accelerating wakefield by optimal beam loading is non-trivial for the non-linear positron-accelerating wake. To approximately flatten the wakefield, we consider a current profile that captures the salient features of the one described in Ref.~\cite{Diederichs:2020}. This is obtained by linearly combining two Gaussian distributions centered in $\zeta_{0,w1} = -11.57\,k_p ^{-1}$ and $\zeta_{0,w2} = -11.3\,k_p ^{-1}$, and with a length of $\sigma_{z,w1} = 0.5\,k_p ^{-1}$ and $\sigma_{z,w2} = 0.2\,k_p ^{-1}$, respectively. The corresponding peak densities are $n_{w1}/n_0 = 260$ and $n_{w2}/n_0 = 44.6$.
The transverse size of the witness bunch is $\sigma_{x,w} = \sigma_{y,w} = 0.029\,k_p^{-1}$, its initial energy is $1\,$GeV, it has no initial energy spread, and the initial normalized emittance is $\epsilon_x = \epsilon_y = 0.1\,k_p^{-1}$ (corresponding to $0.75\,\mu $m\,rad for $n_0=5\times10^{17}$ cm$^{-3}$). The witness bunch is sampled  with $125\times10^6$ macro-particles.

\begin{figure}
	\centering
	\includegraphics[trim={0 0 0 0},clip, width=3.375in]{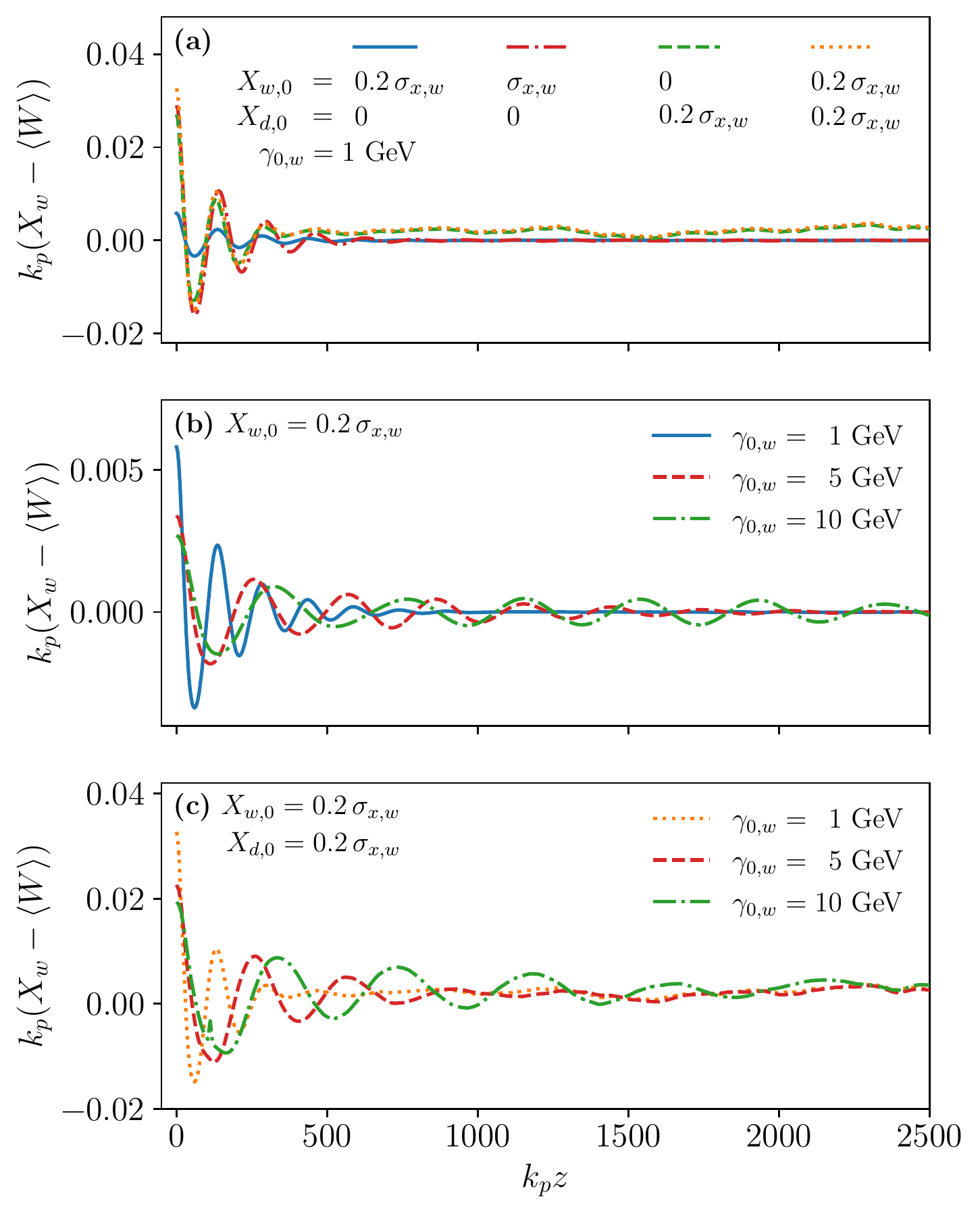}
	\caption{(a) Difference between witness beam centroid $X_w$ and focusing wake centroid $\langle W \rangle$ for different initial witness and/or drive beam offsets as a function of the propagation distance. The witness beam centroid quickly converges to the wake centroid within a few damped oscillations, demonstrating the stability of the scheme versus initial offsets. The cases of a misaligned driver (b) and both a misaligned driver and misaligned witness beam (c) are also shown for higher witness energies of 5 and 10 GeV, respectively. Despite an increased damping length, the evolution is still stable.\label{fig:results}}
\end{figure}

After presenting the basic numerical setting, we now analyze the evolution of the witness beam in the case of various misalignments of the witness beam centroid, the drive beam centroid, or both with respect to the plasma column axis.
The evolution of the witness bunch centroid $X_w$ is shown in Fig.~\ref{fig:results}(a), where we plot the difference between the witness beam centroid $X_{w}$ and the wakefield centroid at the phase location of the witness bunch $\langle W \rangle$. 
In the next paragraphs, different beam configurations are reviewed in detail.

First, the setup of an \textit{on-axis drive beam} and a \textit{misaligned witness beam} is considered. In this configuration, a small initial transverse witness beam offset ($X_{w,0} = 0.2 \,\sigma_{x,w}$, solid blue line) and a larger witness beam offset ($X_{w,0} =  \sigma_{x,w}$, dash-dotted red line) are tested. For both offsets, the witness beam centroid quickly converges to the wake centroid via a strongly damped oscillation. For the small initial transverse beam offset, we also test higher initial witness beam energies, namely 5 GeV (dashed red line) and 10 GeV (dash-dotted green line) as shown in Fig.~\ref{fig:results}(b). To keep the witness beams quasi-matched~\cite{Diederichs:2019}, their transverse rms sizes are reduced $\sigma_{x,w} = \sigma_{y,w} = 0.017\,k_p^{-1}$ and $\sigma_{x,w} = \sigma_{y,w} = 0.0135\,k_p^{-1}$ for 5 GeV and 10 GeV, respectively. The initial offset is kept constant relative to the reduced beam size with $X_{w,0} = 0.2 \,\sigma_{x,w}$. As shown in Fig.~\ref{fig:results}(b), a higher witness beam energy results in a longer damping length. 

Second, the setup of an \textit{misaligned drive beam} and an \textit{on-axis witness beam} is considered. For initial displacements of $X_{d,0} = 0.2\, \sigma_{x,w}$ and $X_{w,0}=0$ (dashed green line in Fig.~\ref{fig:results}(a)), the witness beam centroid again quickly converges to the wake centroid in a damped oscillation. Small deviations of the witness beam centroid from the wake centroid are caused by inaccuracies in the numerical extraction of the wake centroid. Notably, the witness beam centroid stays aligned with the wake centroid, even if the wake centroid is changing due to evolution of the drive beam. 

Finally, the same behaviour is also observed in the most challenging setup of a \textit{misaligned drive beam} and a \textit{misaligned witness beam}, where both the drive and the witness beams are misaligned by $X_{d,0}= 0.2 \,\sigma_{x,w}$ and $X_{w,0} = 0.2 \,\sigma_{x,w}$, respectively (dotted orange line in Fig.~\ref{fig:results}(a)). Hereby, we chose the most challenging case, where both beams are displaced in the same direction, which increases the offset between $X_w$ and $\langle W \rangle$, see Fig.~\ref{fig:scheme}. Due to its lower energy, the witness beam adjusts to the (slowly) evolving wake centroid driven by the 5 GeV drive beam. Thus, we also tested quasi-matched, higher witness beam energies of 5 GeV (dashed red line in Fig.~\ref{fig:results}(c)) and 10 GeV (dash-dotted green line in Fig.~\ref{fig:results}(c)), to verify the stability even in cases where the witness beam is evolving at a slower rate compared to the drive beam. Despite the increased damping length due to the higher witness beam energy, the witness is still able to follow the wake centroid. We also considered cases where the drive and witness beams are offset in orthogonal directions (results not shown), but we did not observe any detrimental coupling between the motion in the two planes.

In all presented cases, the oscillation of the positron witness beam centroid is quickly damped and the witness centroid converges to the wake centroid. The damping is caused by two effects that result in high stability of the witness beam towards transverse displacements: first, the oscillation is damped due to the longitudinally varying focusing fields (see Fig.~\ref{fig:scheme}), which has also been observed in other positron acceleration schemes~\cite{Silva:2021}. This effect is similar to the Balakin-Novokhatsky-Smirnov (BNS) damping mechanism~\cite{Balakin:1983} in conventional accelerators and has been discussed in the context of quasi-linear wakefields in Ref.~\cite{Lehe:PRL:2017}. Second, the non-linear transverse wakefields cause phase mixing within a single slice of the beam that ultimately leads to a damping of the witness beam centroid motion. 

The scaling of the damping due to the non-linear transverse wakefields can be obtained from a simplified theoretical model, described in Appendix~\ref{sec:damp_length}. In the simplified model (see also Ref.~\cite{Diederichs:2019}), we consider a flat beam (i.e., $\sigma_x \gg \sigma_y$) in a step-like confining wakefield, namely $(E_x - B_y)/E_0 = -\alpha \, \text{sgn}(x)$, where $x$ is the transverse coordinate, $\alpha$ is the strength of the wakefield, and $\mathrm{sgn(x)}$ the sign function. The evolution of the drive beam is neglected (i.e., the confining wake is constant), we neglect head-to-tail effects (i.e., we consider damping in a slice at a given longitudinal location along the beam), and we assume the acceleration to be slow (i.e., the particle energy changes significantly over distances longer than the characteristic betatron period). Finally we assume the centroid displacement is small compared to the characteristic beam size, namely $X_{w,0}\ll \sigma_x$.  With these assumptions, the evolution of the beam centroid $X_w$ is found to be a damped oscillation with a damping length $S_{\mathrm{damp}}$ scaling as
\begin{equation}
    k_pS_{\mathrm{damp}} \propto \sqrt{\frac{k_p\sigma_{x,w} \gamma}{\alpha}} \,.
\end{equation}
Notably, and confirmed by the simulations, the damping length does not depend on the initial offset (at least for small displacements). The scaling of the damping length with $\gamma$, $\alpha$, and $\sigma_{x,w}$ is in good agreement with the simulation results.

Since the damping is caused by phase mixing, the damping length is linked to the spread of the betatron frequencies within the bunch. In the case of a Gaussian betatron wavenumber distribution, the damping length is proportional to $S_{\rm damp, Gauss} \propto 1/\sigma_{k_\beta}$, where $\sigma_{k_\beta}$ is the rms spread of betatron wavenumbers. Although the wavenumber distribution within the bunch is not Gaussian in this case, their distribution can be used to quantify which mechanism dominates the damping process between: (i) BNS damping due to the longitudinally varying focusing strength resulting in a $\zeta$-dependent betatron period for particles in different bunch slices, and (ii) non-linear focusing wakefield resulting in an amplitude-dependent betatron period for the particles in a beam slice. Therefore, we examine the distribution of betatron frequencies within the witness bunch. Using the example with the misaligned witness beam with an initial offset of $X_{w,0} = 0.2 \,\sigma_{x,w}$, the betatron frequencies are extracted from 1D test particles simulations by employing a second-order particle pusher and using the initial focusing field of the PIC simulation along $x$ at various slices in the beam. Note that it is possible to obtain the betatron frequencies directly from the PIC simulation, but it is impractical due to the acceleration of the particles and the required data intensity as a short output period is needed for reasonable accuracy. 

\begin{figure}
	\centering
	\includegraphics[trim={0 0 0 0},clip, width=3.375in]{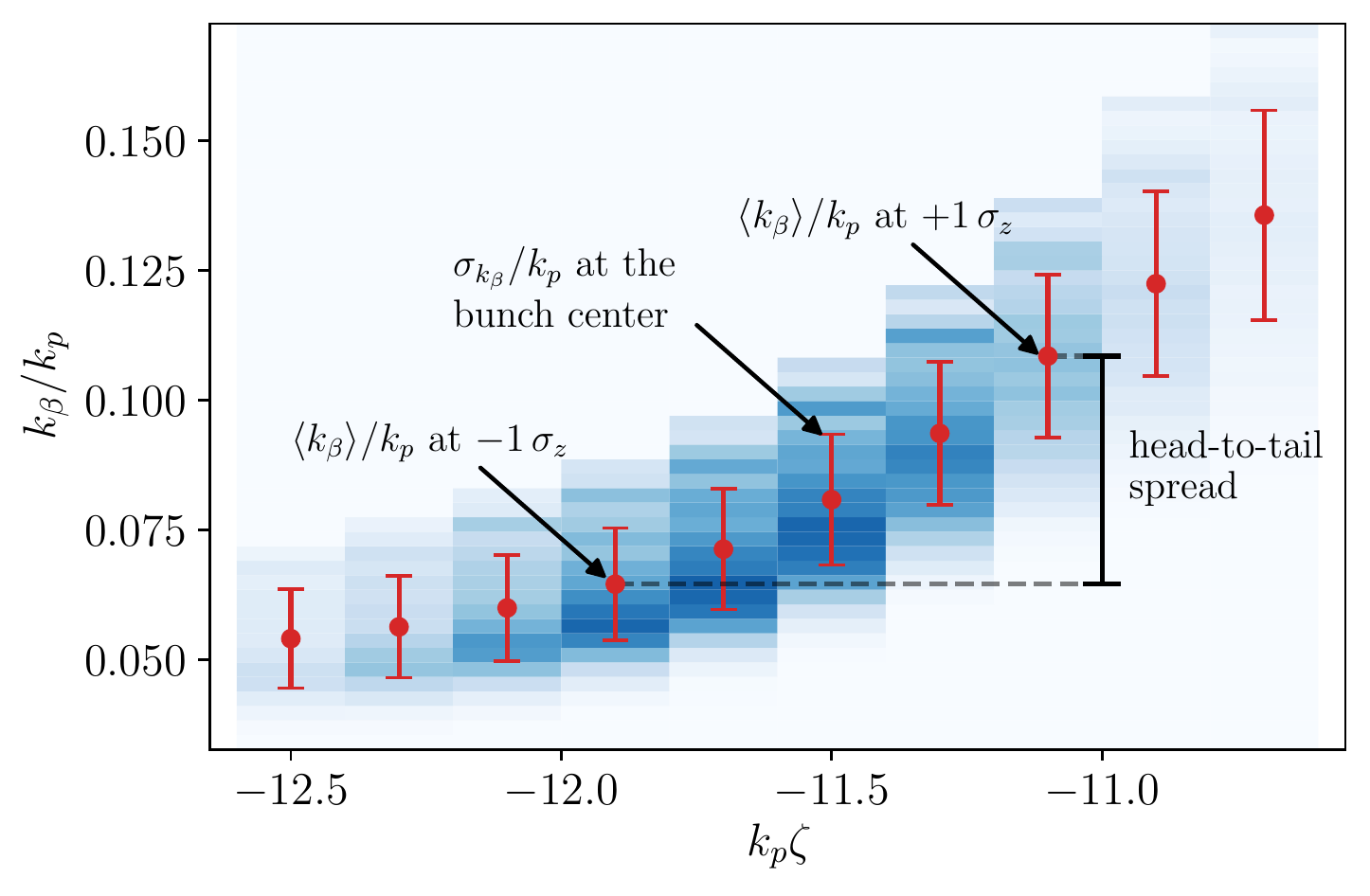}
	\caption{ Distribution of betatron wavenumbers $k_\beta/k_p$ along the co-moving variable $\zeta$. The rms wavenumber spread per slice $\sigma_{k_{\beta}}$ is depicted by the red bars. The head-to-tail wavenumber spread, defined as the difference between the mean wavenumber at the head of the bunch (located at $+1 \sigma_z$) and the tail (located at $-1 \sigma_z$) is depicted by the black bar. The head-to-tail spread is larger than the average of the intra-slice spread.\label{fig:betatron_tune}}
\end{figure}

The distribution of the wavenumbers of the test particles, $k_{\beta}/k_p$, along the bunch is shown in Fig.~\ref{fig:betatron_tune}. The red dots and bars denote the per-slice average $\langle k_{\beta} \rangle$ and the per-slice rms spread $\sigma_{k_{\beta}}$ of the betatron wavenumber, respectively. We define the head and the tail of the bunch as the slices at $\zeta=\sigma_z$ and $\zeta= -\sigma_z$. Then, the head-to-tail wavenumber spread is obtained as the difference between the mean betatron wavenumbers at the head and at the tail, and is visualized in the plot as a black bar. As shown, the head-to-tail wavenumber spread, which is caused by the longitudinal variation of the focusing field, is approximately 4 times larger than the average intra-slice wavenumber spread caused by the transverse non-linearity of the focusing field.
Thus, the variance of the focusing field along the bunch dominates the damping of the positron bunch. Note that the longitudinal variance strongly depends on the beam-loading of the witness bunch.

\section{Effect of misalignment on witness beam quality}
\label{sec:effect_misalignment}

\begin{figure}
	\centering
	\includegraphics[trim={0 0 0 0},clip, width=3.375in]{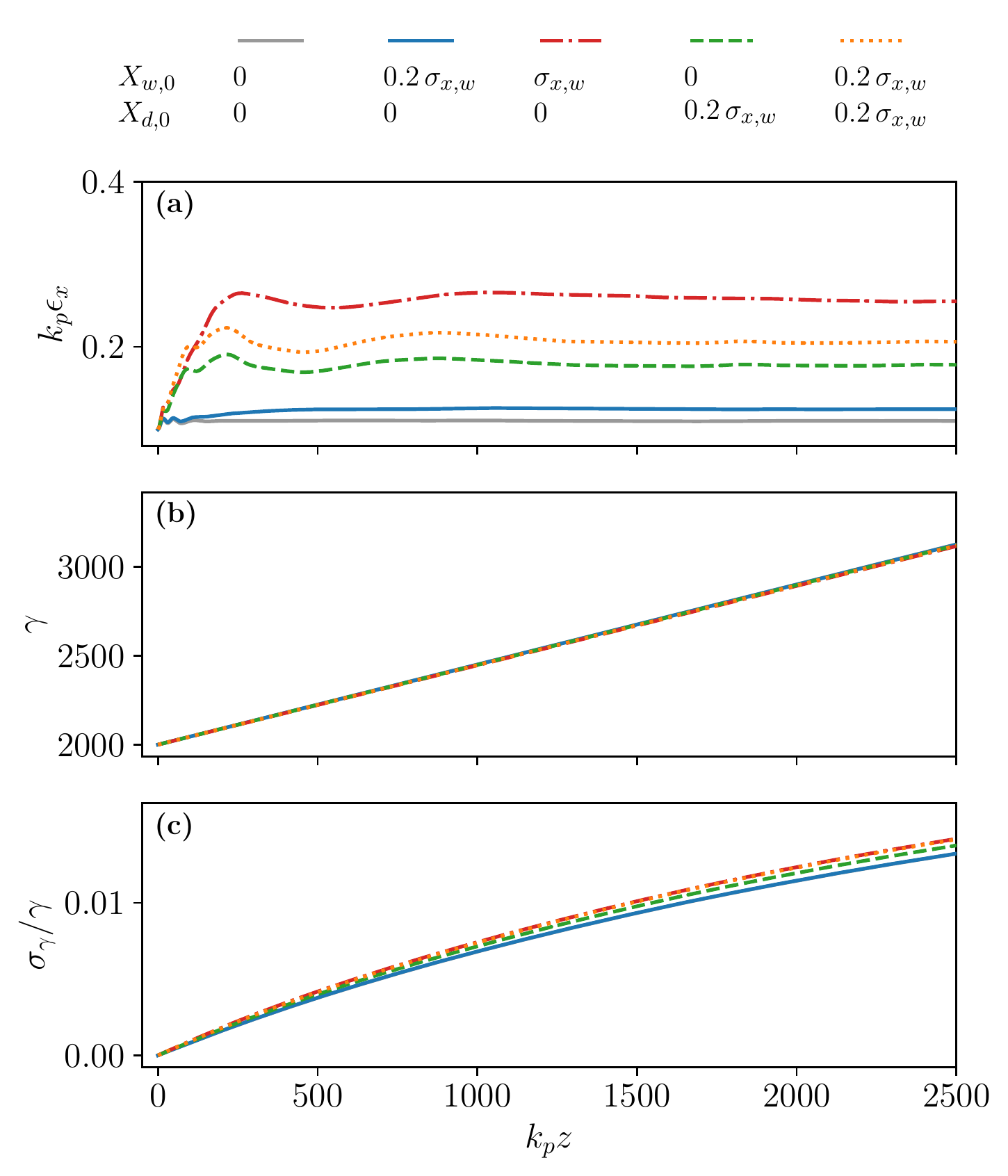}
	\caption{(a)--(c) Emittance, energy gain, and relative energy spread as a function of the propagation distance. Depending on the offset, emittance growth can be observed, which saturates as soon as the beam is aligned with the wake centroid. Only marginal differences in energy gain and energy spread are observed.\label{fig:effect_misalignment}}
\end{figure}

Having discussed the positron stability, we now address the effect of an initial misalignment on the beam quality. For the cases described in Fig.~\ref{fig:results}(a), the effect of the misalignment on the emittance, energy gain, and relative energy spread of the positron witness beam are shown in Fig.~\ref{fig:effect_misalignment}(a)--(c). For a small witness beam offset (solid blue line), the emittance grows only by a few percent in comparison to the aligned case (solid grey line). A large witness beam offset (dash-dotted red line), a drive beam offset (dotted green line), or both (dashed orange line) degrade the emittance due to the larger initial difference between wake centroid and beam centroid. Nevertheless, the emittance growth saturates as soon as the beam centroid converges to the wake centroid. This is also the case for the misaligned drive beam, where the wake centroid is evolving. The energy and the relative energy spread are not sensitive to initial misalignments of the beam centroids, and results differ only marginally from the aligned case. The positron bunch charge is conserved in all studied cases.

To gain further understanding of the effect of an initial offset of the witness beam centroid on the emittance, we use the model for the emittance growth at saturation presented in Ref.~\cite{Diederichs:2019}. In the model, the same general assumptions as in the previous section are made. Then, assuming a small relative initial offset of the witness beam $\Delta_x = X_{w,0}/\sigma_x \lesssim 1$, the emittance growth at saturation is given by
\begin{equation}
\label{eq:em_growth}
\begin{split}
 \frac{\sqrt{\langle x^2\rangle \langle u_x^2\rangle}}{\sigma_x \sigma_{u_x}}  \approx & \bigg\{ \frac{8}{45}\bigg[ \left(1+\frac{4}{\pi}\right)\left(1+\Delta_x^2\right) \\ &+\sqrt{\frac{2}{\pi}}\eta^{-1} \left(2 + 3\Delta_x^2\right) \\ &+\frac{5}{2}\sqrt{\frac{2}{\pi}}\eta \left(1+\frac{\Delta_x^2}{2}\right)+\frac{3}{4}\eta^2 \bigg] \bigg\}^{1/2} \, ,
 \end{split}
\end{equation}
with 
\begin{equation}
    \eta = \frac{\sigma_{u_x}^2}{ k_p\sigma_x \gamma \alpha} \, .
\end{equation}
A detailed derivation of Eq.~\eqref{eq:em_growth} is presented in Appendix~\ref{sec:em_growth}.

We now compare the emittance growth obtained from the model with results from the PIC simulations.
The emittance growth at saturation is estimated using Eq.~\eqref{eq:em_growth} with an energy of $\gamma = 2000$, an initial transverse beam size of $\sigma_{x} = 0.029 \,k_p^{-1}$, an initial emittance of $\epsilon_{0} = 0.1 \,k_p^{-1}$, and a field strength of $\alpha = 0.5$, which is extracted from the PIC simulation with aligned beams with respect to the column axis. For the aligned case with $\Delta_x = 0$, the small offset case with $\Delta_x = 0.2$, and the large offset case with $\Delta_x = 1$ the predicted emittance growth is $6\,\%$, $8\,\%$, and $50\,\%$, respectively. While the emittance growth predicted by the simplified model are in reasonable agreement with the simulation in the aligned and small offset cases, it is significantly underestimated in the large offset case. A reason for the underestimation is that the beamloading effect on the transverse wakefield of a high-charge, strongly misaligned witness beam is not captured by the model. In case of a large offset, the transverse focusing field is asymmetrically altered, thus the assumption of a simple step-like wake in the model is not fulfilled anymore. Therefore, the model provides a reasonable scaling only for small offsets. Additionally, some of the assumptions (flat beam, neglect of head-to-tail effects) made in the model are not fulfilled in the full PIC simulation, so the model should be used for qualitative purposes only.

In summary, the non-linearity of the focusing field transversely provides stability as initial offsets are quickly damped due to the phase mixing of the particles. However, this mechanism comes at a cost, since the phase mixing increases the witness beam size and the emittance.

\section{Conclusion}
\label{sec:conclusion}
In this paper, positron acceleration in a plasma column is shown to be inherently stable:

in case of misalignments between the drive and witness beams with respect to the plasma column, the witness beam centroid is attracted to the centroid of the focusing wake and the witness beam is not susceptible to the hosing instability due to a longitudinally varying and transversely non-linear focusing field in the region of the positron bunch. The initial misalignment is corrected via a damped oscillation, which affects the witness beam quality. A scaling for the damping length of the oscillation and the induced emittance growth is presented. By analysis of the betatron wavenumbers distribution within the bunch, the variation of the focusing field along the bunch is identified to be the dominating effect in the damping. 
To achieve high-quality positron acceleration, the alignment of drive and witness beam with respect to the plasma column is essential, but achievable due to the intrinsic stability, which allows for the implementation of active feedback loops. Plasma columns enable stable, low-emittance, low-energy-spread positron acceleration as required for a linear collider. Initial experiments of this configuration may be performed at beam test facilities, such as FACET-II~\cite{Yakimenko:2019}.

%
%

%
%

\begin{acknowledgments}
We acknowledge the Funding by the Helmholtz Matter and Technologies Accelerator Research and Development Program. This work was supported by the Director, Office of Science, Office of High Energy Physics, of the U.S. Department of Energy, under Contract No. DE-AC02-05CH11231, and used the computational facilities at the National Energy Research Scientific Computing Center (NERSC). We gratefully acknowledge the Gauss Centre for Supercomputing e.V. (www.gauss-centre.eu) for funding this project by providing computing time through the John von Neumann Institute for Computing (NIC) on the GCS Supercomputer JUWELS at J\"ulich Supercomputing Centre (JSC). This research was supported in part through the Maxwell computational resources operated at Deutsches Elektronen-Synchrotron DESY, Hamburg, Germany.

The input scripts for the PIC simulations used in Fig.~\ref{fig:results} and Fig.~\ref{fig:effect_misalignment} are available online~\cite{Diederichs:2022:WBS:dataset}. The data that support the other figures are available upon reasonable request from the corresponding author.
\end{acknowledgments}

\appendix

\section{Damping length of the positron witness beam centroid oscillations}
\label{sec:damp_length}

Here we derive an expression for the characteristic damping length of the positron witness beam centroid oscillations. In the derivation, the following assumptions are made:
\begin{enumerate}
    \item The transverse wakefield close to the axis is approximately step-like and is described by $(E_x - B_y)/E_0 = -\alpha \,\mathrm{sgn}(x)$, with $\mathrm{sgn(x)}$ being the sign function, and the field strength $\alpha > 0$ a constant, respectively.
    \item The drive beam evolution is neglected and the transverse wakefield is constant in time. 
    \item The beam is considered to be flat, i.e., $\sigma_{w,x} \gg \sigma_{w,y}$ and $\sigma_{u_x} \gg \sigma_{u_y}$, with $\sigma_{u_x}$ and $\sigma_{u_y}$ being the transverse rms momentum spreads in $x$- and $y$-direction, respectively.
    \item The acceleration process is adiabatic (i.e., the single particle action is conserved).
    \item A small centroid offset is considered, namely $X_{w,0}\ll \sigma_{x,w}$.
\end{enumerate} 

The damping length depends on the spread in betatron frequencies of the particles in the beam and this, in turns, depends on the beam initial phase space distribution. 
Analytical calculations for the damping length are tractable in case of an initial phase-space distribution that is perfectly matched (in the case without centroid displacement) in the idealized step-like wakefield (i.e., the unperturbed phase-space distribution is a stationary solution of the Vlasov equation for the system). Note that, as explained in Ref.~\cite{Diederichs:2019}, the Gaussian distribution that is used as initial condition in all the simulations presented in this work is not a stationary solution of the Vlasov equation. Hence, in our calculations we will consider an exponential distribution (see below for details on its definition). We verified numerically that the damping length for an exponential distribution is comparable to that of a quasi-matched initial Gaussian distribution with the same rms size (the exponential distribution is a good proxy for a Gaussian distribution). Extension of the results to other types of matched initial distributions is straightforward.

The calculation of the beam centroid evolution requires first determining the centroid motion for a displaced beam with a Kapchinskij-Vladimirskij (KV)~\cite{Kapchinskij:1959} phase-space distribution (KV beamlet, note that an undisplaced KV beam is also a stationary solution of the Vlasov equation, see Sec.~\ref{Sec:AppA1}). Then, the centroid evolution for the whole beam is obtained by considering that a generic matched beam can be decomposed into a sum of KV beamlets (see Sec.~\ref{Sec:AppA2}).

\subsection{Evolution of the beam centroid of a single KV beamlet\label{Sec:AppA1}}

The KV phase-space distribution is defined as~\cite{Kapchinskij:1959}
\begin{equation}
    f_{\rm KV}(x,u_x) \propto \delta \left[H(x, u_x) - H_0\right] \, ,
\end{equation}
were $\delta$ is the Dirac delta distribution, $H(x, u_x)$ is the single particle Hamiltonian for a particle in a step-like wakefield, and it is given by~\cite{Diederichs:2019}
\begin{equation}\label{Hamil}
H(x, u_x) =  \frac{u_x^2}{2\gamma} + \alpha k_p |x| \, ,
\end{equation} 
with $x$, $u_x$, and $\gamma$ being the position, the momentum, and the Lorentz factor of the particle, respectively. Finally, $H_0$ is a parameter that depends on the initial condition and sets the characteristic size of the beamlet. We also introduce the parameter $X_0$, defined as
\begin{equation}
   k_p X_0 = \frac{H_0}{\alpha}\,.
\end{equation}
Since the phase-space distribution depends on the coordinates via the Hamiltonian, then it is, by construction, a stationary solution of the Vlasov equation and so the beamlet is matched (i.e., all the beam moments are constant in time). 
In a KV beamlet all the particles satisfy, at all the times, the condition $H(x, u_x) = H_0$  or, equivalently,
\begin{equation} \label{eq:KV}
    \frac{u_x^2}{2\gamma} + \alpha k_p |x| = \alpha k_p X_0 \,.
\end{equation}
Thus, the particle momentum is a function of the position according to
\begin{equation}
\label{eq:ux}
    u_x = \pm \sqrt{2\gamma\alpha k_p (X_0 - |x|)}\,.
\end{equation}
We see that $|x|\le X_0$, and so $X_0$ represents the maximum  coordinate for a particle in a KV beamlet. The betatron period $L_{\beta}$ of a single particle in the step-like wake is obtained by integrating the equation of motion $dx/ds = u_x/\gamma$, with $u(x)$ given by Eq.~\eqref{eq:ux}, over a closed orbit, yielding
\begin{equation}
    L_{\beta, {\rm KV}} = 2\int_{-X_{0}}^{X_{0}} \sqrt{\frac{\gamma}{2 \alpha k_p (X_{0}- |x|)}} dx\,=8\,\sqrt{\frac{\gamma X_0}{2 \alpha k_p}}\,.
\end{equation}

All the particles in a KV beamlet have the same betatron period that depends on the beamlet size, $X_0$.
The betatron wavenumber is then given by $k_{\beta,KV} = 2\pi / L_{\beta, {\rm KV}}$, or
\begin{equation} \label{eq:kbeta}
    k_{\beta, {\rm KV}}(X_0)\equiv k^0_{\beta} = \frac{\pi}{4} \sqrt{\frac{2\alpha k_p}{\gamma X_0}}\, .
\end{equation}

\begin{figure}
	\centering
	\includegraphics[trim={0 0 0 0},clip, width=3.375in]{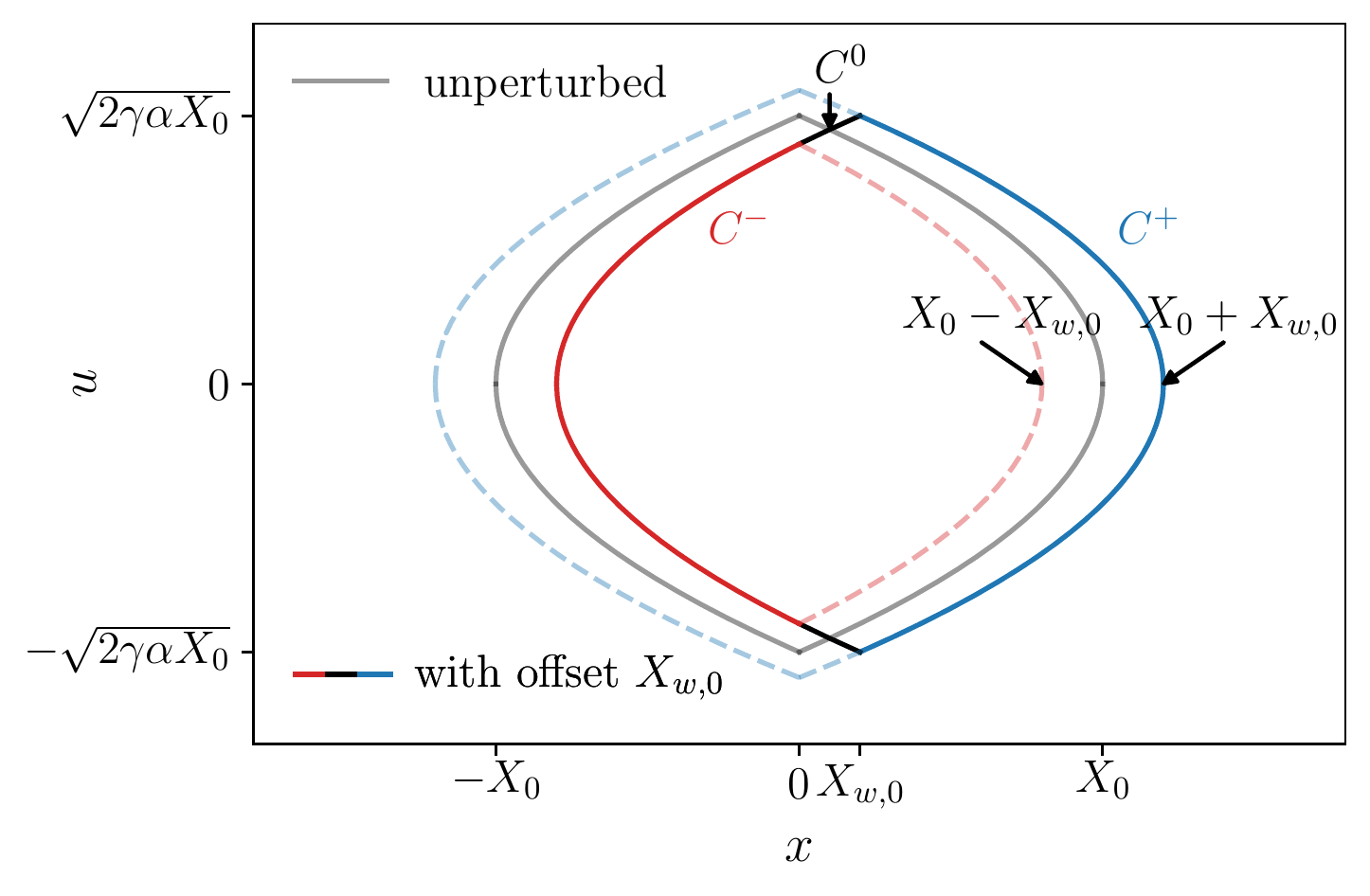}
	\caption{Phase-space of an unperturbed (solid grey line) KV beamlet and of a KV beamlet with an initial offset $X_{w,0}>0$ (solid blue-red-black line). The KV beamlet with an initial offset is not an equilibrium distribution anymore: the particles that are initially on the blue ($C^+$) and red ($C^-$) arch continue their trajectories on the blue and red dashed lines, respectively. Thus, they are effectively part of KV beamlets with sizes $X_0 + X_{w,0}$ (blue) and $X_0 - X_{w,0}$ (red), respectively. The particles that are initially on the black arches ($C^0$) fill trajectories in between the red and blue orbits (filamentation). \label{fig:phasespace_KV_beam}} 
\end{figure}

We now consider a KV beamlet whose centroid is initially displaced by $X_{w,0}$ (in the following, we assume $X_{w,0}>0$ for illustration purposes).  The particles in such beamlet satisfy
\begin{equation}
    \frac{u_x^2}{2\gamma} + \alpha k_p |x-X_{w,0}| = \alpha k_p X_0 \,.
\end{equation}
Notably, a displaced KV beamlet is not an equilibrium distribution anymore, as shown in Fig.~\ref{fig:phasespace_KV_beam}, which displays the phase space of the unperturbed KV beam (grey solid line) and that of the displaced KV beamlet (blue-red-black solid line). The particles in the displaced beamlet can be categorized into three subsections:
first, the particles with $x < 0$ (solid red line, we denote this set of particles with the symbol $C^-$) that satisfy
\begin{equation} \label{eq:C-}
    \frac{u_x^2}{2\gamma} + \alpha k_p |x| = \alpha k_p \left(X_0 - X_{w,0} \right) \,.
\end{equation}
Second, the particles with $x > X_{w,0}$ (solid blue line, $C^+$) that satisfy
\begin{equation} \label{eq:C+}
    \frac{u_x^2}{2\gamma} + \alpha k_p |x| = \alpha k_p \left( X_0 + X_{w,0} \right) \,,
\end{equation}
and third, the remaining particles with $0 < x < X_{w,0}$ (solid black lines, $C^0$). When comparing Eqs.~\eqref{eq:C-} and \eqref{eq:C+} with Eq.~\eqref{eq:KV}, one can see that the particles in $C^-$ and $C^+$ are effectively part of KV beamlets with parameters $X_0 - X_{w,0}$ and $X_0 + X_{w,0}$, respectively. Thus, during evolution, the particles initially in $C^-$ and $C^+$ remain in their respective perturbed KV orbits with sizes $X_0 \mp X_{w,0}$, as indicated by the dashed lines in the corresponding colors in Fig.~\ref{fig:phasespace_KV_beam}. The particles in $C^0$ fill the space area between the red and the blue orbits (filamentation), but the fraction of such particles is small for small centroid displacements. Hence, we see that we can represent a displaced KV beamlet as the sum of two half KV beamlets of different characteristic sizes. From Eq.~\eqref{eq:kbeta} the betatron wavenumbers for the particles in $C^\pm$ are $k_\beta^\pm = k_{\beta, {\rm KV}} (X_0 \pm X_{w,0})$, which reduce, for small offsets $X_{w,0}/X_0 \ll 1$, to
\begin{equation}
    k_\beta^\pm = k_{\beta}^0\left( 1 \mp \frac{1}{2} \frac{X_{w,0}}{X_0}\right) \,.
\end{equation}

To calculate the evolution of the centroid of the full, displaced KV beamlet $X_{w, {\rm KV}}$, the centroids and the fraction of particles in $C^-$, $C^+$, and $C^0$ must be computed.
Starting with the latter, we define $f^-$, $f^+$, and $f^0$ as the fraction of particles in $C^-$, $C^+$, and $C^0$, respectively. We have $f^- + f^+ + f^0 = 1$. 
For small offsets, $X_{w,0}/X_0 \ll 1$, we can neglect the fraction of particles in $C^0$, and we modify $C^+$ by adding the particles in $0 < x < X_{w,0}$ along the dashed blue orbit.
In this case, we have $f^- + f^+ = 1$. The fraction in $C^{\pm}$ is obtained by integrating the corresponding phase-space distribution over the suitable domain. For instance, we have
\begin{equation}
    f^- = \frac{1}{N} \int_{-\infty}^0 dx  \int_{-\infty}^\infty du_x \,\delta\left[\frac{u_x^2}{2\gamma} + \alpha k_p(|x| - X_0 + X_{w,0}) \right] \,,
\end{equation}
with the normalization 
\begin{equation}
     N = \int_{-\infty}^\infty dx  \int_{-\infty}^\infty du_x \,\delta \left(\frac{u_x^2}{2\gamma} + \alpha k_p |x| - \alpha k_p X_0 \right) \, .
\end{equation}
We obtain $f^- = (1/2) \sqrt{1 - X_{w,0}/X_0}$, and assuming $X_{w,0}/X_0 \ll 1$ we have 
\begin{equation} \label{eq:f-}
    f^- \simeq \frac{1}{2} \left(1 - \frac{1}{2} \frac{X_{w,0}}{X_0}\right) \,,
\end{equation}
and so
\begin{equation}  \label{eq:f+}
    f^+ = \frac{1}{2} \left(1 + \frac{1}{2} \frac{X_{w,0}}{X_0}\right) \, .
\end{equation}
The centroid of the KV beamlet at any time is given by
\begin{equation} \label{eq:Xb_raw}
    X_{w, {\rm KV}}(s) = f^- X^-(s) + f^+ X^+(s) \,,
\end{equation}
with $X^{\pm}(s)$ being the centroids of $C^{\pm}$, respectively. Note that since particles in $C^{\pm}$ will rotate clockwise following the respective orbits with betatron periods $k_{\beta}^{\pm}$, we can assume that the corresponding centroids are performing an harmonic motion with the same betatron periods, hence
\begin{equation} \label{eq:Xbpm}
    X^\pm(s) = X^\pm(0) \cos k_\beta^\pm s  \, .
\end{equation}
Here $X_w^{\pm}(0)$ are the initial $C^\pm$ centroids computed by averaging $x$ over the corresponding initial distributions, 
\begin{equation} \label{eq:Xbpm0}
    X^\pm(0) = \pm \frac{2}{3} \left( X_0 \pm X_{w,0}\right)  \, .
\end{equation}
Inserting Eqs.~\eqref{eq:f-}, \eqref{eq:f+}, \eqref{eq:Xbpm}, and \eqref{eq:Xbpm0} into Eq.~\eqref{eq:Xb_raw}, we have
\begin{multline} \label{eq:Xb_centroid}
    X_{w, {\rm KV}}(s) = -\frac{X_0}{3} \left(1-\frac{3}{2} \frac{X_{w,0}}{X_0} \right) \cos k_\beta^- s  \\
    + \frac{X_0}{3} \left(1+\frac{3}{2} \frac{X_{w,0}}{X_0} \right) \cos k_\beta^+ s .
\end{multline}
Keeping only first order terms in $X_{w,0}/X_0$, the expression for $X_{w, {\rm KV}}(s)$ can be further simplified to 
\begin{multline}
\label{eq:exact}
    X_{w, {\rm KV}}(s) = \frac{2}{3} X_0 \sin\left(\frac{k_\beta^0}{2} \frac{X_{w,0}}{X_0} s \right) \sin k_\beta^0 s \\
    + X_{w,0} \cos k_\beta^0 s \, .
\end{multline}
We see that, as a consequence of the beating between the betatron frequencies $k_{\beta}^{\pm}$ associated with the particles in $C^{\pm}$, the centroid of a perturbed KV beamlet performs transverse oscillations with an amplitude that changes periodically in time. The periodicity of the modulation is determined by $X_{w,0}$.  

Finally, we can further simplify Eq.~\eqref{eq:exact} for early times, $(X_{w,0}/X_0) k_{\beta}^0 s \ll 1$, to
 \begin{equation}
 \label{eq:centroid_simplified}
    X_{w, {\rm KV}}(s) \simeq X_{w,0} \left[  \cos k_{\beta}^0 s +  \frac{k_{\beta}^0 s}{3} \sin k_{\beta}^0 s \right] \,.
 \end{equation}

\subsection{Evolution of the centroid for a beam with an exponential phase-space distribution \label{Sec:AppA2}}
We now compute the centroid motion for a displaced beam with an exponential phase-space distribution. The exponential phase-space distribution, including the correct normalization (i.e., such that $\int \int f(x, u_x) dx du_x=1$), is defined as 
\begin{equation}
    f(x, u_x) = \frac{\alpha}{\sqrt{8 \pi \gamma h_0^3}} \exp{ \left[ - \frac{H(x, u_x)}{h_0} \right] } \, ,
\end{equation}
where $h_0$ is a parameter setting the width of the beam. The rms size of the bunch is given by $\sigma_{\mathrm{exp}} \equiv \sqrt{ \langle x^2 \rangle }= \sqrt{2} h_0 / \alpha$. 

A beam with an exponential phase-space distribution can be decomposed into a sum of different KV beamlets of different sizes. From geometric considerations we have that the fraction of particles in a  KV beamlets with sizes between $X_0$ and $X_0 + dX_0$ is given by
\begin{equation}
\label{eq:dN}
    dN(X_0) = \frac{2^{7/4}}{\sqrt{\pi} \sigma_{\mathrm{exp}}^{3/2}} \sqrt{X_0} \exp{ \left( -\sqrt{2}\frac{X_0}{\sigma_{\mathrm{exp}}} \right) } dX_0 \, .
\end{equation}

The motion of the centroid for the whole beam, $X_w$, in case of an initial centroid displacement is calculated by superimposing the centroids of different KV beamlets, Eq.~\eqref{eq:centroid_simplified}, taking into account the weighting given by Eq.~\eqref{eq:dN}, we obtain   
\begin{equation}
\label{eq:KV_centroid}
    X_w (s) = \int_0^{\infty} X_{w, {\rm KV}}(s;X_0) dN(X_0),
\end{equation}
where we explicitly indicated the dependence of $X_{w, {\rm KV}}$ on $X_0$.
After some algebra we obtain the following expression for the evolution of the whole beam centroid 
 \begin{equation}
 \label{eq:full_evolution}
     X_w (s) = X_{w,0} \frac{4}{\pi} \int_0^{\infty} \left[ \cos{(\psi_0 t)} + \frac{\psi_0 t}{3} \sin{(\psi_0 t)} \right] \frac{e^{-\frac{1}{t^2}}}{t^4} \, dt,
 \end{equation}
 with 
 \begin{equation}
 \label{eq:def_psi}
     \psi_0 = \psi_0 (s) = \frac{\pi}{4} \left[ \frac{2 \sqrt{2} \alpha}{k_p \sigma_{\mathrm{exp}} \gamma} \right]^{\frac{1}{2}} k_p s  \,.
\end{equation}
For short propagation distances, $k_p s\ll (4\sqrt{2} k_p\sigma_{\mathrm{exp}} \gamma /  \alpha)^{1/2}/\pi$ (i.e., when $\psi_0 \ll 1$), the solution Eq.~\eqref{eq:full_evolution} can be approximated as
\begin{equation}
    \frac{X_w(s)} { X_{w,0}} \simeq 1 - \frac{\psi_0^2}{3}=1 - \frac{\pi^2 \sqrt{2} \alpha}{24 \sigma_{\mathrm{exp}} \gamma} k_p s^2\,.
\end{equation}

\begin{figure}
	\centering
	\includegraphics[trim={0 0 0 0},clip, width=3.375in]{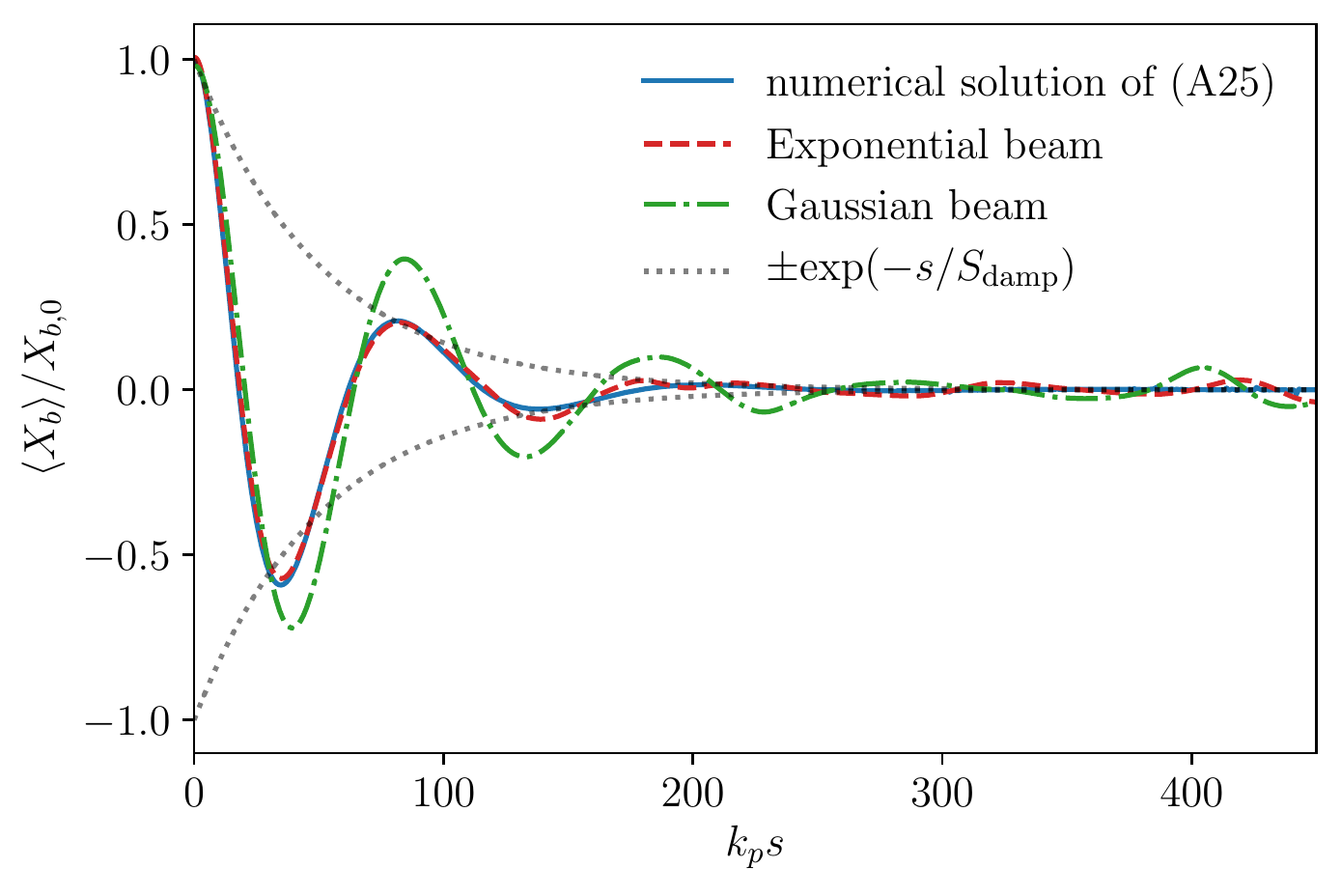}
	\caption{Damping in a step-like transverse focusing field during propagation time $s$. The numerical solution of Eq.~\eqref{eq:full_evolution} agrees well with the reduced modeling of an exponential beam and with an exponential decay assuming the damping length $S_{\mathrm{damp}}$. The damping of a Gaussian beam in the reduced model takes slightly longer but the physical scaling was found to be the same. \label{fig:damping_sdamp}} 
\end{figure}

For longer propagation distances $\psi_0(s)$ grows, and for large enough values of $\psi_0$ the fast oscillating trigonometric functions in Eq.~\eqref{eq:full_evolution} will cause the integral to vanish. Hence, Eq.~\eqref{eq:full_evolution} describes the damping of the centroid oscillations. A numerical study of Eq.~\eqref{eq:full_evolution} shows that this occurs for $\psi_0 \gtrsim 2\pi$. From this, and using Eq.~\eqref{eq:def_psi}, we derive the following expression for the characteristic distance over which damping of the centroid oscillations occurs,
\begin{equation}
\label{eq:s_damp}
   k_p s \gtrsim k_p S_{\mathrm{damp}} = 4 \left[ \frac{\sqrt{2} k_p \sigma_{\mathrm{exp}} \gamma}{\alpha}\right]^{\frac{1}{2}} \, .
\end{equation}

Since the damping length was derived for a beam with an exponential phase-space distribution, while all the simulations in this work use an initial Gaussian phase-space distribution, we verified numerically that the damping length agrees with that of a Gaussian beam to a reasonable level. For the comparison we assumed that $\sigma_{\mathrm{exp}} = \sigma_{x,w}$, with $\sigma_{x,w}$ being the rms size of the Gaussian witness beam. The evolution of the centroid is obtained with test-particles simulations (using a second-order particle pusher) for beams in a step-like wakefield. Results are shown in Fig.~\ref{fig:damping_sdamp}, where we compare the centroid evolution from test particles simulations for an exponential beam (red dashed line), and for a Gaussian beam (green dash-dotted line) to the  numerical solution of Eq.~\eqref{eq:full_evolution} (blue solid line). The physical parameters used in this example are $\alpha = 0.5$, $\gamma=2000$, $\sigma_{x,w} = 0.029\, k_p^{-1}$, $X_{w,0} = 0.005\,k_p^{-1}$, and an emittance of the Gaussian beam of $\epsilon_x = 0.1\,k_p^{-1}$. We find that the solution of Eq.~\eqref{eq:full_evolution} is in excellent  agreement with the simulation results for an exponential beam, and in good agreement with that of a Gaussian beam. We also see that the envelope of the damped centroid oscillations can be well approximated by the exponential $\pm \exp(-s/S_{\mathrm{damp}})$, with $S_{\mathrm{damp}}$ given by Eq.~\eqref{eq:s_damp} (grey dotted lines). 

Note that in the full PIC simulations presented in this work many of the assumptions discussed here are only approximately fulfilled (e.g., round beams are used instead of flat beams, the actual confining wakefield is not an exact step-like function, the drive beam is evolving, etc.) and thus we expect the damping length derived here to be only qualitatively correct.

\section{Emittance growth at saturation with initial offset}
\label{sec:em_growth}

Under the assumptions of the model described in Sec.~\ref{sec:stability_witness}, the emittance growth at saturation is calculated similarly to Refs.~\cite{Benedetti:2017, Diederichs:2019}. The Hamiltonian for a particle in a step-like transverse wakefield, $(E_x - B_y)/E_0 = -\alpha \, \text{sgn}(x)$, is given by
\begin{equation}\label{Hamil}
H(x, u_x) =  \frac{u_x^2}{2\gamma} + \alpha k_p |x| \, .
\end{equation}
The maximum of any particle trajectory $x_{max}$ is obtained by setting $u_x = 0$ for a given initial condition $(x_0, u_{x_0})$, yielding
\begin{equation} \label{xmax}
x_{max} = H(x_0, u_{x_0}) / \alpha \, .
\end{equation}
Using equation (\ref{Hamil}) and (\ref{xmax}), the momentum $u_x$ for any given initial condition is
\begin{equation}
u_x = \pm \sqrt{2\gamma \alpha k_p (x_{max} - |x|)} \, ,
\end{equation}
with $|x| \leq x_{max}$.

Following Ref.~\cite{Benedetti:2017}, the time-average of phase-space moments over a closed particle orbit is given by
\begin{subequations} \label{uxbars}
\begin{align}
\overline{x^2}(x_0, u_{x_0}) &=  \left. \frac{\partial_h \mathcal{P}_2}{\partial_h \mathcal{P}_0} \right\rvert_{h=H(x_0, u_{x_0})=\alpha k_p x_{max} } \\[10pt]
\overline{u_x^2}(x_0, u_{x_0}) &= \left. \frac{\gamma \mathcal{P}_0}{\partial_h \mathcal{P}_0}  \right\rvert_{h=H(x_0, u_{x_0})=\alpha k_p x_{max} }  \, ,
\end{align}
\end{subequations}
with
\begin{equation} \label{pformel}
\mathcal{P}_{\ell}(h) = \int_{0}^{x_{max}} x^{\ell} \left(h - \alpha k_p  |x|\right)^{1/2} dx\, .
\end{equation}
Evaluating Eq.~(\ref{uxbars}) with the given Hamiltonian yields
\begin{subequations}\label{averagepsv}
\begin{align}
\overline{x^2}(x_0, u_{x_0}) &= \frac{8}{15} x_{max}^2 \\[10pt]
\overline{u_x^2}(x_0, u_{x_0}) &= \frac{2}{3} \gamma \alpha  k_p x_{max}  \, .
\end{align}
\end{subequations}

The second-order phase-space moments for the beam are obtained by averaging over the initial Gaussian phase-space distribution. In contrast to Ref.~\cite{Diederichs:2019}, the Gaussian distribution used for the averaging has an initial offset of $X_{w}$:
\begin{subequations} \label{gaussianpsv}
\begin{align}
\begin{split}
\langle x^2 \rangle &= \frac{1}{2\pi \sigma_x \sigma_{u_x}} \int_{-\infty}^{\infty}  \int_{-\infty}^{\infty}  \overline{x^2}(x_0, u_{x_0})  \qquad \\
&\qquad \qquad \times \text{exp}\left( -\frac{(x_0 - X_{w})^2}{2\sigma_x^2} -\frac{u_{x_0}^2}{2\sigma_{u_x}^2}\right)  dx_0 du_{x_0} 
\end{split} \\[10pt]
\begin{split}
\langle u_x^2 \rangle &= \frac{1}{2\pi \sigma_x \sigma_{u_x}}  \int_{-\infty}^{\infty}  \int_{-\infty}^{\infty}  \overline{u_x^2}(x_0, u_{x_0})  \qquad \\
&\qquad \qquad  \times \text{exp}\left( -\frac{(x_0 - X_{w})^2}{2\sigma_x^2} -\frac{u_{x_0}^2}{2\sigma_{u_x}^2}\right)  dx_0 du_{x_0} \, .
\end{split}
\end{align}
\end{subequations}
Solving the integrals in Eq.~(\ref{gaussianpsv}) gives the second-order phase-space moments
\begin{subequations} \label{gaussianpsvfinal}
\begin{align}
\begin{split}
\langle x^2 \rangle =& \frac{8}{15} \Bigg\{ \sigma_x^2 + X_{w}^2+ \frac{ \sigma_{u_x}^2}{\gamma \alpha k_p} \bigg[ \sigma_x e^{-\Delta_x^2/2} \sqrt{\frac{2}{\pi}} \\ &+ X_{w}\,\mathrm{Erf}\left( \frac{\Delta_x}{\sqrt{2}}\right)\bigg] + \frac{3}{4} \frac{\sigma_{u_x}^4}{\gamma^2 \alpha^2} \Bigg\}
\end{split} \\[10pt]
\begin{split}
    \langle u_x^2 \rangle =& \frac{1}{3} \Bigg\{ \sigma_{u_x}^2 + 2 \gamma \alpha k_p \bigg[ \sigma_x e^{-\Delta_x^2/2}\sqrt{\frac{2}{\pi}}  \\ &+ X_{w} \,\mathrm{Erf}\left( \frac{\Delta_x}{\sqrt{2}}\right)  \bigg] \Bigg\} \, ,
\end{split}
\end{align}
\end{subequations}
with $\Delta_x = X_{w}/\sigma_x$ and the error function $\mathrm{Erf}(x) = \frac{2}{\sqrt{\pi}} \int_0^x e^{-t^2}\mathrm{d}t$.
Finally, the relative emittance growth is given by the emittance at saturation $\epsilon_f = \sqrt{\langle x^2 \rangle \langle u_x^2 \rangle} $ divided by the initial emittance $\epsilon_i = \sigma_x \sigma_{u_x}$:

\begin{multline}
\frac{\sqrt{\langle x^2\rangle \langle u_x^2\rangle}}{\sigma_x \sigma_{u_x}} = \bm{\Bigg[ } \frac{8}{45} \bm{\Bigg( }1+\frac{4}{\pi}e^{-\Delta_x} + \Delta_x^2 \\ 
+ 4 e^{-\Delta_x/2} \sqrt{\frac{2}{\pi}} \Delta_x \, \mathrm{Erf}\left( \frac{\Delta_x}{\sqrt{2}}\right) + 2 \Delta_x \, \mathrm{Erf}\left( \frac{\Delta_x}{\sqrt{2}}\right)^2 \\ 
+ 2\eta^{-1} \left\{ \left[1+\Delta_x^2\right] \left[\sqrt{\frac{2}{\pi}} e^{-\Delta_x^2/2} + \Delta_x  \,\mathrm{Erf}\left( \frac{\Delta_x}{\sqrt{2}}\right) \right] \right\} \\   +\frac{5}{2}\eta \left[\sqrt{\frac{2}{\pi}}e^{-\Delta_x^2/2} + \Delta_x \,\mathrm{Erf}\left( \frac{\Delta_x}{\sqrt{2}}\right) \right]    +\frac{3}{4}\eta^2 \bm{\Bigg) \Bigg] }^{1/2} \, ,
\end{multline}

with 
\begin{equation}
    \eta = \frac{\sigma_{u_x}^2}{k_p \sigma_x \gamma \alpha} \, .
\end{equation}
A second-order Taylor expansion for $\Delta_x \ll 1$ yields

\begin{equation}
\begin{split}
 \frac{\sqrt{\langle x^2\rangle \langle u_x^2\rangle}}{\sigma_x \sigma_{u_x}}  \approx & \bigg\{ \frac{8}{45}\bigg[ \left(1+\frac{4}{\pi}\right)\left(1+\Delta_x^2\right) \\ &+\sqrt{\frac{2}{\pi}}\eta^{-1} \left(2 + 3\Delta_x^2\right) \\ &+\frac{5}{2}\sqrt{\frac{2}{\pi}}\eta \left(1+\frac{\Delta_x^2}{2}\right)+\frac{3}{4}\eta^2 \bigg] \bigg\}^{1/2} \, .
 \end{split}
\end{equation} 
The Taylor expansion is accurate for $\Delta_x \lesssim 1$ and underestimates the emittance growth at saturation by less than $1\,\%$ and $5\,\%$ for $\Delta_x = 0.5$ and  $\Delta_x = 1$, respectively.

\bibliography{Bibliography}

\end{document}